\documentclass[conference,10pt]{IEEEtran}
\usepackage{amsmath,amssymb,amsfonts}
\usepackage{algorithmic}
\usepackage{graphicx}
\usepackage{textcomp}
\usepackage{xcolor}
\usepackage[caption=false,font=footnotesize]{subfig}

\def\BibTeX{{\rm B\kern-.05em{\sc i\kern-.025em b}\kern-.08em
    T\kern-.1667em\lower.7ex\hbox{E}\kern-.125emX}}
\usepackage{amssymb,amsmath}
\usepackage{xspace}
\usepackage{bm}
\usepackage{bbm}
\usepackage{mathtools}

\let\originalleft\left
\let\originalright\right
\renewcommand{\left}{\mathopen{}\mathclose\bgroup\originalleft}
\renewcommand{\right}{\aftergroup\egroup\originalright}

\newcommand{\mat}[1]{{\mathbf{#1}}} % matrices
\renewcommand{\vec}[1]{{\mathbf{#1}}} % vectors
\newcommand{\parens}[1]{{\left(#1\right)}\xspace}
\newcommand{\brackets}[1]{{\left[#1\right]}\xspace}

\newcommand{\logtwo}[1]{\ensuremath{\mathrm{log}_{2}\parens{#1}}}

\newcommand{\sinp}[1]{{\mathrm{sin}\parens{#1}}}

\newcommand{\distexp}[1]{\mathrm{Exponential}\parens{#1}\xspace}

\newcommand{\distgamma}[2]{\mathrm{Gamma}\parens{#1,#2}\xspace}

\newcommand{\ev}[1]{\mathbb{E}\brackets{#1}}

\def\vx{{\vec{x}}}

\def\mPhi{{\mat{\Phi}}}

\newcommand{\rmc}{\mathrm{c}}

\newcommand{\rmf}{\mathrm{f}}

\newcommand{\rml}{\mathrm{l}}
\newcommand{\rmm}{\mathrm{m}}

\newcommand{\rmM}{\mathrm{M}}

\newcommand{\rmS}{\mathrm{S}}

\newcommand{\figref}[1]{Fig.~\ref{#1}}

\begin{document}
\title{Adaptive Cell Range Expansion in\\Multi-Band UAV Communication Networks}

\author{
\IEEEauthorblockN{Xinsong Feng and Ian P.~Roberts}
\IEEEauthorblockA{Wireless Lab, Department of Electrical and Computer Engineering, UCLA, Los Angeles, CA, USA\\
Email: \{xsfeng, ianroberts\}@ucla.edu}%
}
\maketitle

\begin{abstract}
This paper leverages stochastic geometry to model, analyze, and optimize multi-band unmanned aerial vehicle (UAV) communication networks operating across low-frequency and millimeter-wave (mmWave) bands. 
We introduce a novel approach to modeling mmWave antenna gain in such networks, which allows us to better capture and account for interference in our analysis and optimization. 
We then propose a simple yet effective user-UAV association policy, which strategically biases users towards mmWave UAVs to take advantage of lower interference and wider bandwidths compared to low-frequency UAVs. 
Under this scheme, we analytically derive the corresponding association probability, coverage probability, and spectral efficiency.
We conclude by assessing our proposed association policy through simulation and analysis, demonstrating its effectiveness based on coverage probability and per-user data rates, as well as the alignment between analytical and simulation results.
\end{abstract}

\section{Introduction}

Unmanned aerial vehicles (UAVs) are widely used in various fields including aerial photography, disaster relief, and wireless communications \cite{mozaffari2019tutorial}, thanks to their affordability and mobility. 
In wireless communications, UAV base stations (BSs) can provide temporary connectivity, expand coverage, and improve network reliability and efficiency by supplementing or replacing traditional ground BSs \cite{akdeniz2014Millimeter}. 
Meanwhile, \textit{multi-band} networks, like 5G, leverage both low- and high-frequency spectrum to balance coverage and capacity needs \cite{chakareski2019energy}. 
This attractive paradigm motivates the potential of multi-band UAV-based networks, but their analysis and optimization remain open problems, marking the focus of this work.

As with all networks, successful design and deployment of UAV networks relies on thorough performance analyses.
In this context, stochastic geometry provides a tractable mathematical framework for modeling the randomness of real-world network deployments and has been employed widely in analyzing traditional cellular networks \cite{andrews2011tractable}, but its use has been limited in UAV networks.
For instance, in analyzing a single-band UAV network, the authors of \cite{raja2022Coverage} used a Poisson point process (PPP) to model the location of UAVs and then derived the downlink coverage probability. 
The authors in \cite{shi2022Modeling} used a binomial point process to study mmWave UAV networks, using a 3D blockage model and a 3D sectorized antenna model to capture air-to-ground propagation at mmWave frequencies.
The work of \cite{yang2023stochastic} studied mobile UAVs serving ground users modeled by a Poisson cluster process and derived the success probability under hybrid automatic repeat request.

Like traditional ground BSs, UAVs operating at mmWave frequencies rely on antenna arrays and beamforming to overcome high-frequency propagation loss. 
However, accurately modeling beamforming gain within a stochastic geometry framework remains a significant challenge.
Previous work has used simplified antenna gain models, such as flat-top, sinc, and cosine patterns \cite{yu2016Coverage, yu2017Coverage}, or assumed uniform beam steering distributions \cite{shi2022Modeling}. 
While tractable, these approaches compromise accuracy, motivating the need for more precise antenna gain modeling, especially in networks with interference.

Strictly speaking, cell association and coverage optimization in dense multi-band networks is a challenging task that involves balancing signal strength, interference, and BS load across multiple frequency bands. 
A simple yet effective approach to this problem is so-called \textit{cell range expansion} (CRE), which improves coverage and performance by tuning only a few key system parameters.
For example, an adaptive CRE scheme was introduced in \cite{nakazawa2017enhanced} for heterogeneous networks, using transmit power control to enhance capacity and throughput.  
In other work \cite{zhang2017Energy}, the authors used bias factors to manage offloading between macro and small cells through environment-specific optimization in two-tier heterogeneous networks.
Similarly, the authors in \cite{hattab2018Ratebased} presented a rate-based user association rule with a bias factor to improve long-term data rates, but relied on hand-tuning the bias factor, limiting its adaptability.
It remains unclear how to create CRE schemes which are effective yet simple, deployment-friendly, and based purely on network statistics/parameters, not real-time conditions.

In this paper, we introduce a stochastic geometry-based analytical framework for multi-band UAV networks and derive the corresponding association probability, coverage probability, and spectral efficiency.
Core to our approach is a novel approximation of mmWave UAV antenna gain, which improves interference modeling compared to existing approaches \cite{shi2022Modeling}.
Based on this, we introduce an adaptive CRE scheme that leverages network statistics rather than real-time factors, enhancing its practicality.
Extensive simulations confirm the accuracy of our analytical results and demonstrate that the proposed CRE scheme significantly improves network coverage and per-user data rates.

\begin{figure}[t]
    \centering
    \includegraphics[width=3.4in]{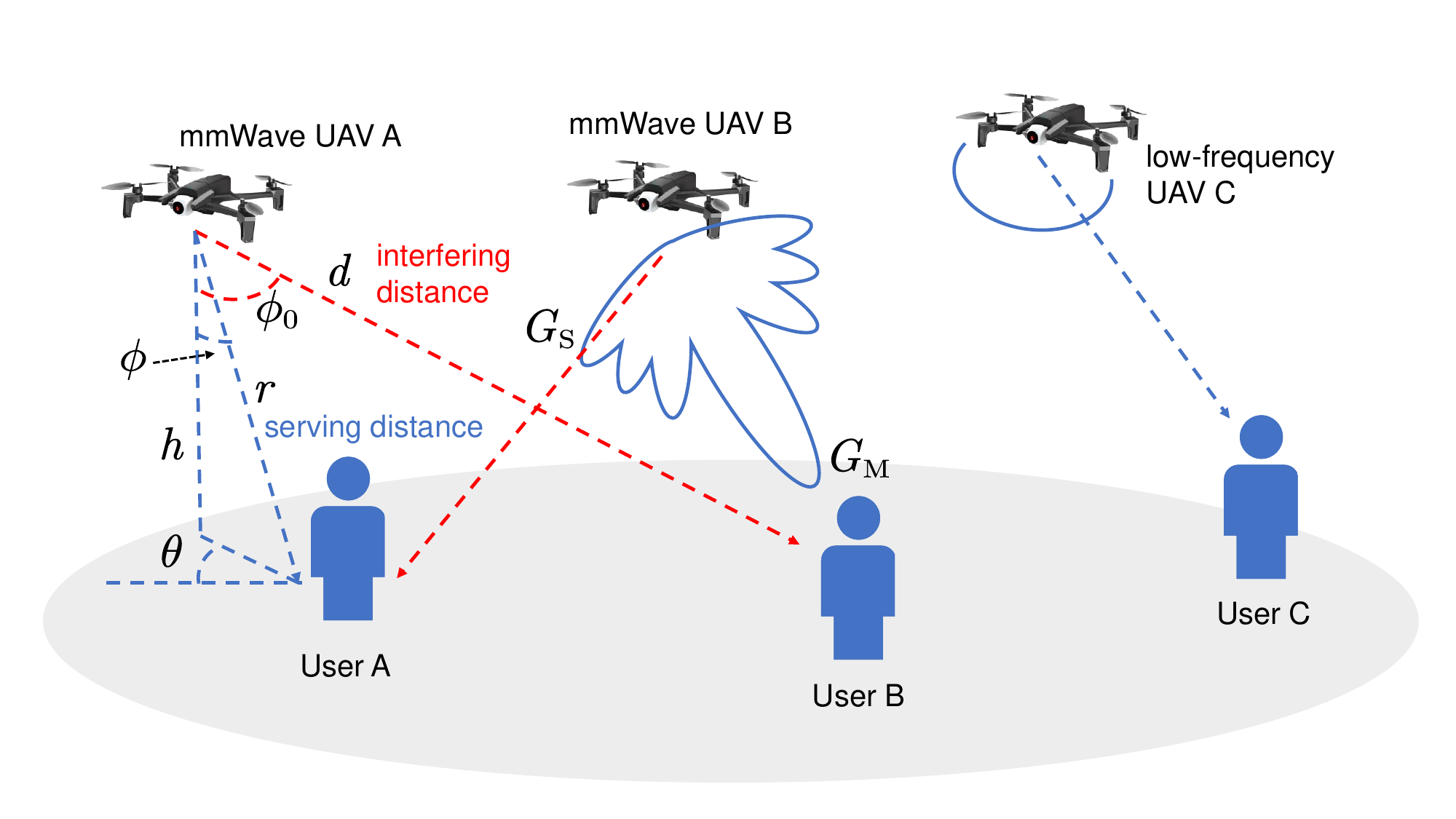}
    \caption{UAVs at height $h$, users on the ground. Users associate with either low-frequency or mmWave UAVs, with interference from non-serving UAVs.}
    \label{fig:topology}
\end{figure}

\section{System Model}

Illustrated in \figref{fig:topology}, this work considers a multi-band UAV network serving downlink to ground users.
The network includes two types of UAV BSs: one operating at a relatively low frequency (e.g., 1~GHz) and the other at a mmWave frequency (e.g., 30~GHz).
We assume all UAVs hover at a height of $h$ meters, with the planar coordinates of the $i$-th UAV denoted by $\vx_i$.
Let $\mPhi_{\mathrm{lf}}$ and $\mPhi_{\mathrm{m}}$ denote the sets of low-frequency and mmWave UAVs, respectively.
The coordinates of the low-frequency UAVs in $\mPhi_\mathrm{lf}$ follow a homogeneous PPP \cite{chiu2013stochastic} with density $\lambda_{\mathrm{lf}}$.
Analogously, $\mPhi_\rmm$ follows a homogeneous PPP with density $\lambda_\rmm$.
Ground users are modeled by a PPP with density $\lambda \gg \lambda_{\rmm}+\lambda_{\rml\rmf}$, ensuring each UAV has at least one associated user with high probability. 
Within either frequency band, each user associates with the UAV providing the highest average received signal strength.

We assume each low-frequency UAV employs a single omnidirectional antenna with unit gain in all directions.
In contrast, each mmWave UAV is equipped with an antenna array and employs beamforming to overcome mmWave path loss.
Specifically, we consider a square uniform planar array (UPA) of $N$ antennas with half-wavelength antenna spacing. 
Adopting the sectorized UPA gain model in \cite{venugopal2016Devicetodevice}, we model the antenna gain pattern through four parameters: azimuth and elevation half-power beamwidths $(\Delta_\theta, \Delta_\phi)$ and the gains of the main and side lobes $G_{\rmM}, G_{\rmS}$. 
We assume that each mmWave UAV electronically steers the center of its main lobe directly toward its serving user to deliver maximal gain $G_\rmM$. 
All antenna arrays are down-tilted toward the ground, with azimuth $\theta\in [0,2\pi)$ and elevation $\phi \in [0,\frac{\pi}{2})$ defined as in \figref{fig:topology}; note that $\phi=0$ corresponds to directly toward the ground and $\phi=\frac{\pi}{2}$ corresponds to toward the horizon.
Ground users have a single omnidirectional antenna with unit gain.

\begin{table}[t] 
    \centering
    \caption{Uniform Planar Antenna Array Parameters \cite{venugopal2016Devicetodevice}}
    \begin{tabular}{|c|c|}
     \hline Number of antennas    & $N$  \\
     \hline Half-power beamwidth $\Delta_\theta = \Delta_\phi$ & $\sqrt{3/N}$ \\
     \hline Main-lobe gain $G_\rmM$  & $N$ \\
     \hline Side-lobe gain $G_\rmS$  & $\frac{\sqrt{N}-\frac{\sqrt{3}}{2\pi}N \sinp{\frac{\sqrt{3}}{2\sqrt{N}}}}{\sqrt{N}-\frac{\sqrt{3}}{2\pi} \sinp{\frac{\sqrt{3}}{2\sqrt{N}}}}$ \\
     \hline
    \end{tabular}
    
    \label{tab:antenna}
\end{table}

We assume both low-frequency and mmWave signals undergo free-space path loss and small-scale fading. 
The inverse path loss at frequency $f_c$ over a distance $d$ is modeled by $K d^{-\alpha}$, where $K=(\frac{\rmc}{4 \pi f_\rmc})^2$ and $\alpha$ are specific to each band.
Note that only signals in the same frequency band cause interference to one another.
Thus, the received signal-to-interference-plus-noise ratio (SINR) at a target user in either band becomes 
\begin{align}
    \mathrm{SINR}_{\mathrm{lf}}&=\frac{P_{\mathrm{lf}} g_{\mathrm{lf}}K_{\rml\rmf} r^{-\alpha_\mathrm{lf}}}{\sigma^2_{\mathrm{lf}}+I_{\mathrm{lf}}}, \label{eq:SINR_lf} \\ 
    \mathrm{SINR}_{\rmm}&=\frac{P_{\rmm} G_\rmM g_{\rmm} K_\rmm r^{-\alpha_\rmm}}{\sigma^2_{\rmm}+I_{\rmm}},\label{eq:SINR_m}
\end{align}
where $P_{\mathrm{lf}}$ and $P_{\rmm}$ denote the corresponding UAV transmit powers, $\sigma_{\mathrm{lf}}^{2}$ and $\sigma_{\rmm}^2$ are the ground user noise powers, and
$I_{\mathrm{lf}}=\sum_{\vx_i \in \mPhi_{\mathrm{lf}}\setminus \{\vx_{0}\}}P_{\mathrm{lf}}g_{\mathrm{lf}}K_{\rml\rmf} d_{i}^{-\alpha_\mathrm{lf}}$ and $I_{\rmm}=\sum_{\vx_i \in \mPhi_{\rmm} \setminus \{\vx_{0}\}} P_{\rmm}G_{\vx_i} g_{\rmm}K_\rmm d_{i}^{-\alpha_\rmm}$ 
are the aggregate interference terms, where $d_i$ denotes the distance between the $i$-th UAV and the target user.
We assume Rayleigh fading for the low-frequency band and Nakagami-$m$ fading for the mmWave band \cite{shi2022Modeling}. 
Correspondingly, $g_{\mathrm{lf}} \sim \distexp{1}$ and $g_{\rmm} \sim \distgamma{m}{1/m}$. 

\section{Network Performance Analysis}
With our network model laid forth, we now derive expressions to analyze and subsequently optimize network performance via a novel association policy.
We begin by deriving the distance distribution between a UAV and its serving user.

\subsection{Serving Distance Distribution}
The \textit{serving distance} between the typical user at the origin and the nearest UAV in each band, denoted by $R_t$ where $t \in \{\mathrm{lf, m}\}$, follows a probability density function (PDF) given by
\begin{align}\label{eq:dist distribution}
f_{R_{t}}(r) = 2\pi \lambda_{t} r \exp( -\pi \lambda_{t} (r^2-h^2)), \quad r \geq h.
\end{align}
This can be derived by considering UAV locations as a homogeneous PPP with density $\lambda_t$ and applying the nearest-neighbor distance distribution by PPPs \cite{andrews2011tractable}.

\subsection{Antenna Gain of Interfering UAVs} \label{sec:antenna pattern}
The severity of interference inflicted onto a ground user by interfering (non-serving) UAVs plays a decisive role in overall network performance, and characterizing such demands accurate modeling of the antenna gain of mmWave UAVs.
To this end, we derive an expression for the antenna gain of interfering UAVs by leveraging the serving distance distribution in \eqref{eq:dist distribution}.

As illustrated in Fig.~\ref{fig:topology}, consider mmWave UAVs A and B serving their respective users A and B at the same time and frequency.
UAV A inflicts interference onto user B when serving its associated user A.
This interference depends on the antenna gain of UAV A toward user B when steering its beam toward user A.
Let $(\theta,\phi)$ be the azimuth and elevation and $r$ be the distance from UAV A to its associated user A.
Let $(\theta_0,\phi_0)$ be the azimuth and elevation and $d$ be the distance from UAV A to the interfered user B.
Users are associated with UAVs based on average received signal strength, meaning that user B must be closer to UAV B than to UAV A, under our assumed model.

Denoting the location of UAV A by $\vx$, its antenna gain toward user B can be expressed as
\begin{equation} \label{eq:antenna pattern}
G_{\vx}=
\begin{cases}
G_\rmM & \mathrm{w.p.~} p_{\rmM}=p_\theta \cdot p_\phi \\
G_\rmS & \mathrm{w.p.~} p_{\rmS}=1-p_\theta \cdot p_\phi
\end{cases},
\end{equation}
where $p_\theta$ and $p_\phi$ are the probabilities that user B falls within the azimuth and elevation beamwidths, respectively, of UAV A's beam steered toward user A.
The azimuth angles $\theta$ and $\theta_0$ are uniformly distributed across $[0,2\pi)$, since UAVs and users follow independent homogeneous PPPs.
Consequently, the probability that the azimuth $\theta_0$ from UAV A to user B is within the main lobe of the beam steered by UAV A toward $(\theta,\phi)$ is $p_{\theta} = \frac{\Delta_\theta}{2\pi}$.
Arriving at $p_\phi$ proves to be a little more involved, which we tackle as follows.

Assuming that $\Delta_\phi$ is reasonably small and the mmWave UAV density $\lambda_\mathrm{m}$ is not extremely high, it is with high probability that $\phi_0-\frac{\Delta_\phi}{2} \geq 0$ and $\phi_0 +\frac{\Delta_\phi}{2} \leq \frac{\pi}{2}$, which we assume henceforth. 
Conditioning on $d$, the distance between UAV A and user B, the probability that $\phi_0$ falls in the main lobe of the beam steered by UAV A toward $(\theta,\phi)$ is
\small
\begin{align}
p_{\phi} 
&= \mathbb{P}\left( \phi_0 -\frac{\Delta_\phi}{2} \leq \phi\leq \phi_0 +\frac{\Delta_\phi}{2} \right)\\ 
&\approx \mathbb{P}\left(\cos\left(\phi_0 -\frac{\Delta_\phi}{2}\right)\geq \cos \phi \geq \cos\left( \phi_0 +\frac{\Delta_\phi}{2} \right)\right) \\
&= \mathbb{P}\left( \frac{h}{\cos\left(\phi_0 -\frac{\Delta_\phi}{2}\right)} \leq r \leq \frac{h}{\cos\left( \phi_0 +\frac{\Delta_\phi}{2} \right)} \right) \\
&= F_{R_{\mathrm{m}}} \left(\frac{h}{\cos\left( \phi_0 +\frac{\Delta_\phi}{2} \right)}\right)- F_{R_{\mathrm{m}}}\left(\frac{h}{\cos\left(\phi_0 -\frac{\Delta_\phi}{2}\right)}\right) \label{eq:prob_a} \\
&\overset{\textrm{(a)}}{=} \Delta_\phi \cdot f_{R_{\mathrm{m}}}\left(\frac{h}{\cos \phi_{c}} \right) \cdot \frac{{h \sin \phi_{c}}}{\cos^{2}\phi_{c}}  \label{eq:prob_b} \\
&\overset{\textrm{(b)}}{\approx} \Delta_\phi \cdot f_{R_{\mathrm{m}}}(d) \cdot \frac{d\sqrt{ d^{2}-h^{2} }}{h} \label{eq:prob_c} \\
&\overset{\textrm{(c)}}{=} 2\pi \lambda_{\mathrm{m}} \Delta_\phi  \exp(-\pi \lambda_{\mathrm{m}} (d^{2}-h^{2})) \cdot \frac{d^{2}\sqrt{ d^{2}-h^{2} }}{h}, \label{eq:prob_d}
\end{align}
\normalsize
where \textrm{(a)} follows from the mean value theorem for some 
% $\phi_c \in \left[ \phi_0-\frac{\Delta_\phi}{2},\phi_0+\frac{\Delta_\phi}{2} \right]$,
$\phi_c \in [\phi_0-\frac{\Delta_\phi}{2},\phi_0+\frac{\Delta_\phi}{2}]$,
\textrm{(b)} follows from the assumption of a narrow beamwidth $\Delta_\phi$
% \xs{[We just use the mean value of the interval, so maybe we can say it's because $\frac{Delta_phi}{2}$ is quite small]} 
and the fact that $\cos \phi_0 = \frac{h}{d}$, and (c) follows from the derivation of serving distance in \eqref{eq:dist distribution}.
With $p_\phi$ closely approximated by \eqref{eq:prob_d}, the antenna gain expression in \eqref{eq:antenna pattern} will be used in statistically analyzing network performance under the proposed CRE scheme introduced next.

\subsection{Cell Range Expansion Scheme and Association Probability}
\begin{figure}[tb]
    \centering
    \includegraphics[width=3.4in]{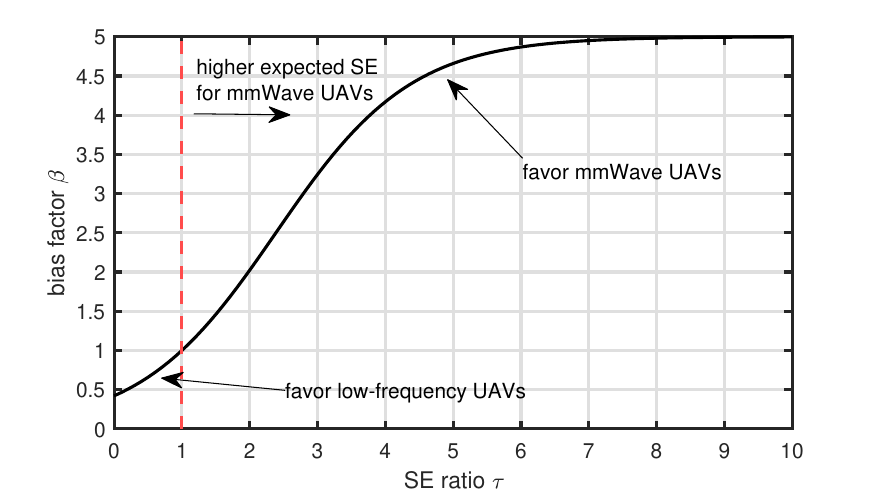}
    \caption{An example bias function with $\beta_0=5, \alpha=1$, and $\zeta=1$. }
    \label{fig:bias}
\end{figure}

Users can associate with either a low-frequency or mmWave UAV based on average received signal power.
A common policy employs biased received power, where a bias factor $\beta > 0$ adjusts the preference for mmWave UAVs. 
Let $\overline{S}_{\mathrm{lf}}$ and $\overline{S}_{\mathrm{m}}$ represent the average received power from the nearest low-frequency and mmWave UAVs, respectively. 
The policy is:
\begin{align}
    \begin{cases}
    \mathsf{associate~with~mmWave~UAV}, & \mathrm{if} \ \beta \overline{S}_{\mathrm{m}}>\overline{S}_{\mathrm{lf}}, \\
    \mathsf{associate~with~low\textsf{-}frequency~UAV}, & \mathrm{else}, 
    % \mathbb{P}(\beta \overline{S}_{\mathrm{m}} \leq \overline{S}_{\mathrm{lf}}) \\
    \end{cases}
\end{align}

Prior work often hand-tunes $\beta$ to optimize network performance \cite{zhang2017Energy}. 
In contrast, we define $\beta$ in closed form based on network statistics and system parameters.
We first introduce the SE ratio $\tau$ as
\begin{equation}
    \tau = \frac{\ev{\logtwo{1+\mathrm{SINR}_{\rmm}}}}{\ev{\logtwo{1+\mathrm{SINR}_{\mathrm{lf}}}}},
\end{equation}
where $\mathrm{SINR}_{\mathrm{lf}}$ and $\mathrm{SINR}_{\rmm}$ are given in \eqref{eq:SINR_lf} and \eqref{eq:SINR_m}, respectively.
Then, we define our proposed bias factor $\beta$ as
\begin{equation}\label{eq:beta}
    \beta= \zeta \cdot \frac{\beta_{0}}{1+(\beta_{0}-1)\exp(\alpha(1-\tau))},
\end{equation}
where the standardization term $\zeta$ is given by
\begin{equation}\label{eq:zeta}
    \zeta = \frac{P_\mathrm{lf}K_\mathrm{lf}\ev{r_\mathrm{lf}^{-\alpha_\mathrm{lf}}}}{P_{\mathrm{m}} G_{\rmM}K_\rmm\mathbb{E}[r_{\mathrm{m}}^{-\alpha_{\mathrm{m}}}]}.
\end{equation}
Here, $\beta_{0}$ represents the maximum value of $\beta$, and $\alpha$ controls the growth rate, taking a sigmoid shape. 
$\zeta$ standardizes the comparison between low-frequency and mmWave signals. 

An example bias factor $\beta$ is shown in Fig.~\ref{fig:bias}.
When $\tau=1$, then $\beta=1$ and there is no bias given to either frequency band.
When $\tau > 1$, then $\beta > 1$ and mmWave UAVs are favored.
As $\tau$ increases, $\beta$ saturates to prevent overloading mmWave UAVs, even if their expected SE exceeds that of low-frequency UAVs.
When $\tau < 1$, the bias shifts toward low-frequency UAVs.
Even as $\tau \to 0$, however, $\beta$ is lower-bounded by about $0.5$, meaning the low-frequency signal strength $\overline{S}_{\mathrm{lf}}$ is artificially inflated by at most a factor of around two, relative to $\overline{S}_{\mathrm{m}}$.
This prevents overloading low-frequency UAVs when $\tau$ is extremely low.
As evidenced by \eqref{eq:beta} and \eqref{eq:zeta}, calculating $\beta$ only depends on system parameters and network statistics, namely the expectations $\mathbb{E}[r_{\mathrm{lf}}^{-\alpha_{\mathrm{lf}}}],\mathbb{E}[r_{\mathrm{m}}^{-\alpha_{\mathrm{m}}}]$ and the expected SE ratio $\tau$, the latter of which we derive shortly.

Let us now derive the association probabilities under our proposed scheme.
For a fixed bias factor $\beta$, the probability that a given user associates with a mmWave UAV is
\begin{align}
\mathcal{A}_{\text{m}}&=\mathbb{E}_{r_{\text{m}}}[\mathbb{P}(\beta \overline{S_{\text{m}}}>\overline{S_{\text{lf}}})] \\
&=\mathbb{E}_{r_{\text{m}}}\left[ \mathbb{P}\left( \beta P_{\text{m}}G_{M}K_{\mathrm{lf}} r_{\mathrm{m}}^{-\alpha_{\mathrm{m}}}  >P_{\text{lf}} K_{\mathrm{lf}} r^{-\alpha_{\mathrm{lf}}} \right) \right]\\
&=\mathbb{E}_{r_{\text{m}}}\left[ \mathbb{P}\left( r_{\text{lf}} >  \eta^{\frac{1}{\alpha_{\mathrm{lf}}}}  r_{\text{m}}^{\alpha_{\mathrm{m}}/\alpha_{\mathrm{lf}}} \right) \right]\\
&=\int _{h} ^{\infty }     \left(1-F_{R_{\text{lf}}}\left( \eta^{\frac{1}{\alpha_{\mathrm{lf}}}}  r^{\alpha_{\mathrm{m}}/\alpha_{\mathrm{lf}}} \right)\right)   \cdot f_{R_{\text{m}}}(r)   \,\mathrm{d}r,
\end{align}
where $\eta= \frac{P_{\mathrm{lf}} K_{\mathrm{lf}}}{\beta P_{\mathrm{m}} G_{M} K_{\mathrm{m}}}$.
The association probability for low-frequency UAVs is then $\mathcal{A}_{\mathrm{lf}}=1-\mathcal{A}_\rmm$.

\subsection{Coverage Probability}
Coverage probability is the probability that a user's SINR exceeds a given threshold, also known as the complementary cumulative distribution function of SINR. 
Using the association probability from before, the  coverage probability of the UAV network for some threshold $\gamma$ can be expressed as,
\begin{equation}
\mathcal{P}_{C}(\gamma) =
\mathcal{P}_{C_{\mathrm{lf}}}(\gamma) \cdot \mathcal{A}_{\mathrm{lf}} + 
\mathcal{P}_{C_{\mathrm{m}}}(\gamma) \cdot \mathcal{A}_{\mathrm{m}},
\end{equation}
where $\mathcal{P}_{C_{\mathrm{lf}}}$ and $\mathcal{P}_{C_{\mathrm{m}}}$ represent the independent coverage probability for low-frequency and mmWave, respectively. 
These can be derived as follows.
% \noindent \textit{Low-frequency UAV coverage probability:}
The low-frequency UAV coverage probability is found as
\begin{align}
\mathcal{P} _{C_{\mathrm{lf}}}\left( \gamma \right) 
&=\mathbb{E}[\mathbb{P} (\mathrm{SINR}_{\mathrm{lf}}>\gamma )\\
&=\mathbb{E}\left[ \mathbb{P} \left( \frac{P_{\mathrm{lf}} g_{\mathrm{lf}}K_\mathrm{lf} r^{-\alpha_\mathrm{lf}}}{\sigma _{\mathrm{lf}}^{2}+I_{\mathrm{lf}}}>\gamma \right) \right] \\
&=\mathbb{E}\left[ \mathbb{P} \left( g_{\mathrm{lf}}>u_{1}(r) \cdot \left( \sigma _{\mathrm{lf}}^{2}+I_{\mathrm{lf}} \right) \right) \right] \\
&\overset{\textrm{(a)}}{=}\mathbb{E}\left[ \exp \left( -u_{1}(r) \cdot \left( \sigma _{\mathrm{lf}}^{2}+I_{\mathrm{lf}} \right) \right) \right] \\
&=\int_h^{\infty}\exp \left( -\sigma_{\mathrm{lf}}^{2} u_{1}(r) \right) \cdot \mathcal{L} _{\mathrm{I}_{\mathrm{lf}}}\left( u_{1}(r) \right) \cdot f_{R_{\mathrm{lf}}}\left( r \right)    \,\mathrm{d}r, \nonumber
\end{align}
where $u_1(r)=\frac{\gamma}{P_{\mathrm{lf}}K_\mathrm{lf} r^{-\alpha_\mathrm{lf}}}$ and \textrm{(a)} follows from $g_{\text{lf}} \sim \distexp{1}$. 
% \noindent \textit{mmWave UAV coverage probability:}
The mmWave UAV coverage probability can be found as
\begin{align}
    \mathcal{P}_{C_{\mathrm{m}}}\left( \gamma \right) 
    &= \mathbb{E} \left[\mathbb{P} \left(\mathrm{SINR}_{\mathrm{m}} > \gamma \right) \right] \\
    &= \mathbb{E} \left[ \mathbb{P} \left( \frac{P_{\mathrm{m}} G_\mathrm{M} g_{\mathrm{m}} K_{\mathrm{m}} r^{-\alpha_{\mathrm{m}}}}{\sigma_{\mathrm{m}}^{2} + I_{\mathrm{m}}} > \gamma \right) \right] \\
    &= \mathbb{E} \left[ \mathbb{P} \left( g_{\mathrm{m}} > \frac{\gamma \left( \sigma_{\mathrm{m}}^{2} + I_{\mathrm{m}} \right)}{P_{\mathrm{m}} G_\mathrm{M} K_{\mathrm{m}} r^{-\alpha_{\mathrm{m}}}} \right) \right].
\end{align}

Since $g_\rmm$ follows a Gamma distribution, the result in \eqref{eq:C-m} on the following page is obtained, where $u_2(r)=\frac{\gamma m}{P_{\mathrm{m}} G_{\rmM} K_\rmm r^{-\alpha_\rmm}}$.
We will derive the Laplace transformations of the aggregate interference for both the low-frequency and mmWave bands in Section \ref{sec: Laplace}.

\begin{figure*}[htbp]
\small
\begin{equation}\label{eq:C-m}
\begin{aligned}
\mathcal{P}_{C_{\mathrm{m}}}\left( \gamma \right) 
    &= \mathbb{E}_{I_{\mathrm{m}}, r} \left[ \sum_{k=0}^{m-1} \frac{1}{k!} \left( u_2\left( r \right) \left( \sigma_{\mathrm{m}}^{2} + I_{\mathrm{m}} \right) \right)^k \right. 
	\times \exp \left( -u_2\left( r \right) \left( \sigma_{\mathrm{m}}^{2} + I_{\mathrm{m}} \right) \right) \Bigg] \\
    &= \mathbb{E}_r \left[ \sum_{k=0}^{m-1} \frac{\left( u_2\left( r \right) \right)^k}{k!} \sum_{i=0}^k \binom{k}{i} \left( \sigma_{\mathrm{m}}^{2} \right)^{k-i} \exp \left( -\sigma_{\mathrm{m}}^{2} \cdot u_2\left( r \right) \right) \right.  \times \mathbb{E}_{I_{\mathrm{m}}} \left[ I_{\mathrm{m}}^i \cdot \exp \left( -v I_{\mathrm{m}} \right) \right] \Bigg] \\
    &= \int_h^{\infty} \Bigg[ \sum_{k=0}^{m-1} \frac{\left( u_2\left( r \right) \right)^k}{k!} \sum_{i=0}^k \binom{k}{i} \left( \sigma_{\mathrm{m}}^{2} \right)^{k-i} \exp \left( -\sigma_{\mathrm{m}}^{2} \cdot u_2\left( r \right) \right) \times \left( -1 \right)^i \left[ \frac{\partial^i}{\partial s^i} \mathcal{L}_{\mathrm{m}}\left( s \mid r \right) \right] \bigg|_{s = u_2\left( r \right)} \Bigg] f_{R_{\mathrm{m}}}\left( r \right) \, \mathrm{d}r,
\end{aligned}
\end{equation}
\end{figure*}
\normalsize

\subsection{Spectral Efficiency}

Now, we will investigate the SE of the UAV network.
Assuming each user attains their maximum achievable SE, we derive the average SE for low-frequency and mmWave UAVs separately. 
For a typical user, the expected SE of low-frequency UAVs can be computed as
\small
\begin{align}
\mathrm{SE} _{\mathrm{lf}}
&=\mathbb{E} \left[ \log_{2} \left( 1+\mathrm{SINR}_{\mathrm{lf}} \right) \right] \\
&=\int_h^{\infty} \mathbb{E} \left[ \log_{2} \left( 1+\frac{P_{\mathrm{lf}} g_{\mathrm{lf}} K_{\mathrm{lf}} r^{-\alpha_{\mathrm{lf}}}}{\sigma _{\mathrm{lf}}^{2}+I_{\mathrm{lf}}} \right) \right] f_{R_{\mathrm{lf}}}(r) \, \mathrm{d}r \\
&=\int_h^{\infty} \int_0^{\infty} \mathbb{P} \left( \log_{2} \left( 1+\frac{P_{\mathrm{lf}} g_{\mathrm{lf}} K_{\mathrm{lf}} r^{-\alpha_{\mathrm{lf}}}}{\sigma _{\mathrm{lf}}^{2}+I_{\mathrm{lf}}} \right) > t \right) \nonumber \\
&\qquad \times f_{R_{\mathrm{lf}}}(r) \, \mathrm{d}t \, \mathrm{d}r \\
&=\int_h^{\infty} \int_0^{\infty} \mathbb{P} \left( g_{\mathrm{lf}} > v_{1}(r,t) \left( \sigma _{\mathrm{lf}}^{2}+I_{\mathrm{lf}} \right) \right) f_{R_{\mathrm{lf}}}(r) \, \mathrm{d}t \, \mathrm{d}r \\
&=\int_h^{\infty} \int_0^{\infty} \exp \left( - \sigma _{\mathrm{lf}}^{2} v_{1}(r,t) \right) \mathcal{L}_{I_{\mathrm{lf}}}\left( v_{1}(r,t) \right) f_{R_{\mathrm{lf}}}(r) \, \mathrm{d}t \, \mathrm{d}r.
\end{align}
\normalsize
The expected SE of mmWave UAVs is then found to be
\small%
\begin{align}
\mathrm{SE}_{\mathrm{m}}
&=\mathbb{E}[\log_{2}\left( 1+\mathrm{SINR}_{\mathrm{m}} \right)] \\
&=\frac{1}{\ln 2}\int_h^{\infty} \mathbb{E} \left[ \ln \left( 1+\frac{g_{\mathrm{m}}}{{{\left( \sigma _{\mathrm{m}}^{2}+I_{\mathrm{m}} \right)}/{\left( P_{\mathrm{m}} G_\mathrm{M} K_{\mathrm{m}} r^{-\alpha_{\mathrm{m}}}  \right)}}} \right) \right] \nonumber \\ 
&\qquad \times f_{R_{\mathrm{m}}}\left( r \right) \,\mathrm{d}r \\
&\overset{\textrm{(a)}}{=}\frac{1}{\ln 2}\int_h^{\infty} \int_0^{\infty} \mathcal{K}\left( z \right) \cdot \mathcal{L}_{I_{\mathrm{m}}}\left( u_{2}(r,z) \right) \cdot \exp\left( - \sigma_{\mathrm{m}}^{2} u_{2}(r,z) \right) \nonumber \\
&\qquad \times f_{R_{\mathrm{m}}}\left( r \right)\,\mathrm{d}z\,\mathrm{d}r,
\end{align}
\normalsize%
where $v_{1}(r,t)=\frac{2^{t}-1}{P_{\mathrm{lf}} K_{\mathrm{lf}} r^{-\alpha_{\mathrm{lf}}}}$, $v_{2}(r,z) = \frac{mz}{P_{\mathrm{m}} G_{\mathrm{M}} K_{\mathrm{m}} r^{-\alpha_{\mathrm{m}}}}$, and $\mathcal{K}(z)=\frac{1}{z} -\frac{1}{z(1+z)^{2}}$.
Here, $\textrm{(a)}$ is based on \cite[Part C, Section 2]{hamdi2007useful}.
Then, the expected SE of the UAV network can be computed by $\mathrm{SE} = \mathcal{A}_{\mathrm{m}} \cdot \mathrm{SE}_{\rmm} + \mathcal{A}_{\mathrm{lf}} \cdot\mathrm{SE}_{\mathrm{lf}}$.

\subsection{Laplace Transformation of Interference} \label{sec: Laplace}
We now derive the conditional Laplace transformation of the aggregate interference. 
The Laplace transformations of the total interference encountered by target users in the low-frequency and mmWave bands, conditioned on a distance $r$ to its serving UAV, are given as follows. 
For the low-frequency band, we have

\small
\begin{align}
\mathcal{L} _{I_{\mathrm{lf}}}\left( s|r \right) 
&=\mathbb{E}_{I_{\mathrm{lf}}}\left[ \exp \left( -sI_{\mathrm{lf}} \right) \right] \nonumber \\
&=\mathbb{E}_{\mathbf{x}_{i}, g_{\mathrm{lf}}^{(i)}}\left[ \prod_{\mathbf{x}_i\in {{\mPhi _{\mathrm{lf}}}/{\{ \mathbf{x}_0 \}}}}{\exp \left( -s P_{\mathrm{lf}} g_{\mathrm{lf}}^{\left( i \right)}K_{\mathrm{lf}} d_{i}^{-\alpha_{\mathrm{lf}}} \right)} \right] \nonumber \\
&\overset{\textrm{(a)}}{=}\mathbb{E}_{\mathbf{x}_{i}}\left[ \prod_{\mathbf{x}_i\in {{\mPhi _{\mathrm{lf}}}/{\{ \mathbf{x}_0 \}}}}{\frac{1}{1+s P_{\mathrm{lf}} K_{\mathrm{lf}} d_{i}^{-\alpha_{\mathrm{lf}}}  }} \right] \nonumber \\
&\overset{\textrm{(b)}}{=}\exp \left( -2\pi \lambda _{\mathrm{lf}}\int_r^{\infty}{\left( 1-\frac{1}{1+s P_{\mathrm{lf}} K_{\mathrm{lf}} z^{-\alpha_{\mathrm{lf}}}} \right) \cdot z}\,\mathrm{d}z \right) \nonumber \\
&=\exp \left( -2\pi \lambda _{\mathrm{lf}}\int_r^{\infty}{\frac{z}{1+s^{-1} P_{\mathrm{lf}}^{-1} K_{\mathrm{lf}}^{-1} z^{\alpha_{\mathrm{lf}}}}}\,\mathrm{d}z \right),
\end{align}
\normalsize
where $\textrm{(a)}$ follows from the moment generating function of $g_{\text{lf}}$ and $\textrm{(b)}$ follows from the probability generating function of a PPP, which is $\mathbb{E}\left[ \prod_{\mathbf{x} \in \mPhi} f(\mathbf{x})=\exp\left( -\lambda \int _{\mathbb{R}^2}(1-f(\mathbf{x})) \, \mathrm{d}x \right) \right]$.
And for the mmWave band, we have
\small
\begin{align}
\mathcal{L}_{I_{\mathrm{m}}}\left( s|r\right) 
&= \mathbb{E}_{I_{\mathrm{m}}}\left[ \exp\left( -s I_{\mathrm{m}} \right) \right] \nonumber \\
&= \mathbb{E}\left[ \prod_{\mathbf{x}_i \in {{\mPhi _{\mathrm{m}}}/\{ \mathbf{x}_0 \}}} \exp\left( -s P_{\mathrm{m}} G_{\mathbf{x}_i} g_{\mathrm{m}}^{\left( i \right)} K_{\mathrm{m}} d_{i}^{-\alpha_{\mathrm{m}}} \right) \right] \nonumber \\
&\overset{\textrm{(a)}}{=} \mathbb{E}_{\mathbf{x}_{i}, G_{\mathbf{x}_{i}}}\left[ \prod_{\mathbf{x}_i \in {{\mPhi _{\mathrm{m}}}/\{ \mathbf{x}_0 \}}} \left( 1 + \frac{s P_{\mathrm{m}} G_{\mathbf{x}_i} K_{\mathrm{m}} d_{i}^{-\alpha_{\mathrm{m}}}}{m} \right)^{-m} \right] \nonumber \\
&\overset{\textrm{(b)}}{=} \exp\left( -2\pi \lambda_{\mathrm{m}} \int_r^{\infty} \left( 1 - \right. \right. \nonumber \\
&\qquad \left. \left. \mathbb{E}_{G_{\mathbf{x}}}\left[\left( 1 + \frac{s P_{\mathrm{m}} G_{\mathbf{x}} K_{\mathrm{m}} z^{-\alpha_{\mathrm{m}}}}{m} \right)^{-m} \right] \right) z \, \mathrm{d}z \right), \label{eq:expect}
\end{align}
\normalsize
where $\textrm{(a)}$ follows from the moment generating function of $g_{\text{m}}$ and $\textrm{(b)}$ follows from the probability generating function of a PPP.
Note that, with the distribution of $G_{\mathbf{x}_i}$ given in \eqref{eq:antenna pattern}, the expectation in \eqref{eq:expect} can be computed directly.
In turn, we are able to characterize coverage probability and SE by evaluating the derived analytical expressions.

\section{Simulation Results}

\begin{figure*}[!t]
    \centering
    \subfloat[Coverage probability vs. SINR threshold. ]{\includegraphics[width=\linewidth,height=0.2\textheight,keepaspectratio]{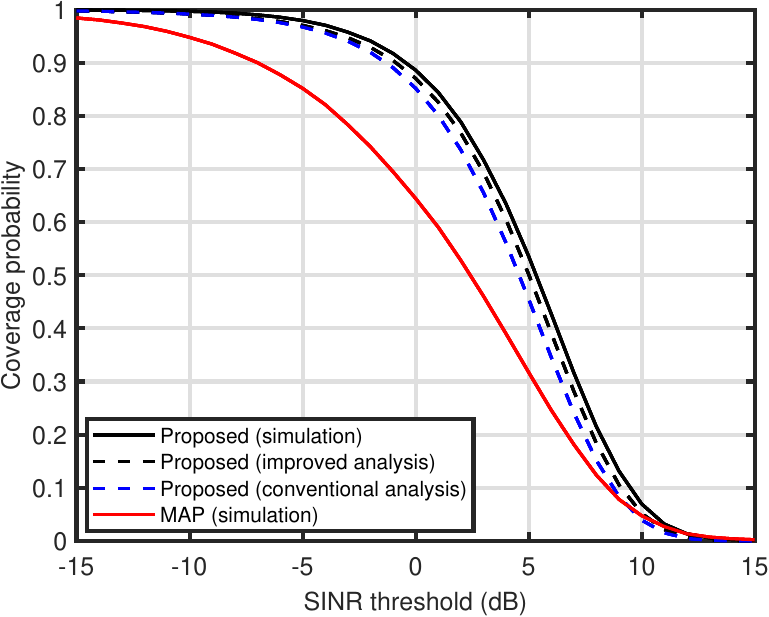} \label{fig:C-SINR}}
    \hfill
    \subfloat[Average per-user data rate vs. mmWave-to-low-frequency UAV density ratio.]{\includegraphics[width=\linewidth,height=0.2\textheight,keepaspectratio]{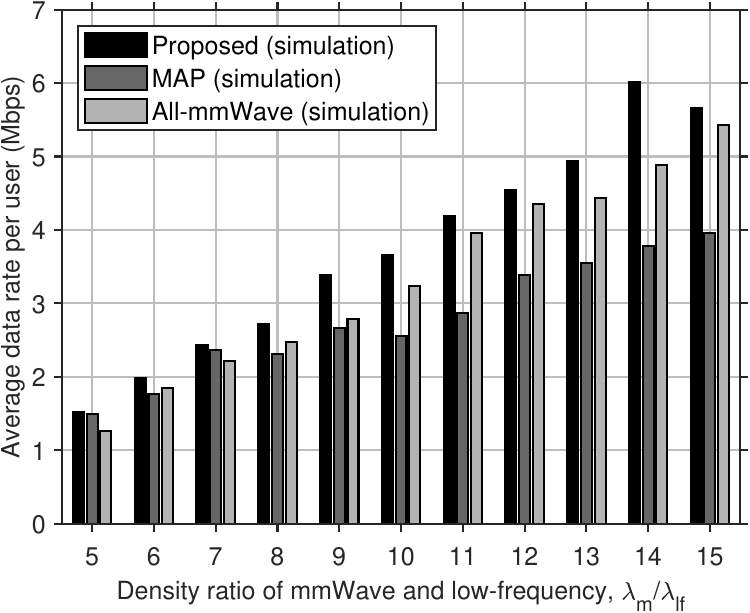} \label{fig:R-ratio}}
    \hfill
    \subfloat[Spectral efficiency vs. the number of mmWave antennas. ]{\includegraphics[width=\linewidth,height=0.2\textheight,keepaspectratio]{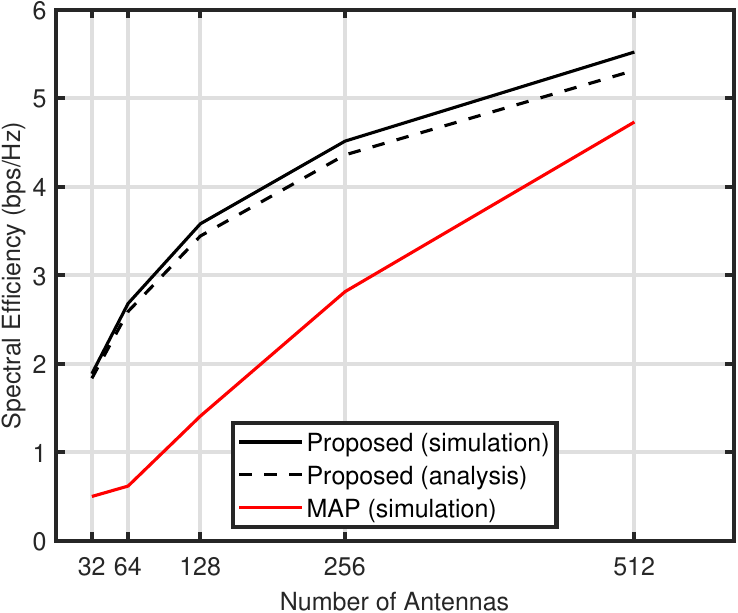} \label{fig:SE-Nt}}
    \caption{Comparison of coverage probability, per-user data rate, and spectral efficiency in multi-band UAV networks under our proposed CRE scheme.}
\end{figure*}

This section validates our analysis and evaluates the proposed CRE scheme.
The experimental parameters are as follows: carrier frequencies $f_{\mathrm{lf}}=2$~GHz and $ f_\rmm=60$~GHz; 
bandwidths $W_\mathrm{lf}=20$~MHz and $W_\rmm=600$~MHz; 
transmit powers $P_\rmm=40$~dBm and $P_{\mathrm{lf}}= 30$~dBm; 
UAV height $h=50$~m; 
UAV densities $\lambda_{\mathrm{lf}}=10$~UAVs/km\textsuperscript{2} and $\lambda_{\mathrm{m}}=500$~UAVs/km\textsuperscript{2}; 
user density $\lambda=5 \times 10^{4}$~users/km\textsuperscript{2};
noise powers $\sigma_{\mathrm{lf}}^{2}=-91$~dBm, $\sigma_{\mathrm{m}}^{2}=-76$~dBm; $N=64$ antennas;
bias factor parameters $\beta_{0}=5$ and $\alpha=5$; path loss exponents $\alpha_\mathrm{lf}=2.5$ and $\alpha_\rmm=3$.

Fig.~\ref{fig:C-SINR} depicts coverage probability versus SINR threshold for four cases. 
The solid black line is empirical coverage probability under our proposed CRE scheme, while the dashed black line is its analytical counterpart using our derived antenna gain distribution.
The dashed blue line uses a simplified antenna gain model assuming uniformly distributed elevation angles \cite{shi2022Modeling}, and the solid red line represents the conventional MAP association policy.
The proposed CRE scheme clearly improves coverage probability compared to the MAP policy, as it accounts for interference by encouraging users to associate with mmWave UAVs.
For instance, under an SINR threshold of $0$~dB, the proposed scheme increases the coverage probability from $65$\% to nearly $90$\%.

In Fig.~\ref{fig:R-ratio}, we vary the mmWave-to-low-frequency UAV density ratio, and the bias factor $\beta$ is tuned accordingly using \eqref{eq:beta}, with $\tau$ and $\zeta$ computed from our derived expressions.
Simulations verify that the proposed CRE scheme significantly improves per-user data rates compared to a conventional MAP policy, especially as mmWave UAV density increases.
However, once the mmWave UAV density reaches a certain threshold, the benefits of offloading become less apparent.
Overall, Fig.~\ref{fig:R-ratio} demonstrates the practicality and effectiveness of our closed-form bias factor $\beta$, which adapts to network statistics and eliminates the need for real-time tuning.

Finally, Fig.~\ref{fig:SE-Nt} depicts the SE versus the numer of mmWave antennas. 
As the number of antenna increases, SE improves due to beamforming, which enhances signal strength and reduces interference.
This increases mmWave SE, attracting more users to mmWave and narrowing the performance gap between the proposed CRE scheme and MAP.

\section{Conclusion}
Using stochastic geometry, this paper has analyzed and optimized multi-band UAV networks comprised of low-frequency and mmWave UAV-mounted BSs. 
Our novel approach to deriving the antenna gain distribution better captures interference under beamforming, providing a more reliable measure of network performance. 
Combined with our proposed CRE scheme, which increases coverage probability and data rates by biasing users toward mmWave UAVs with lower interference and wider bandwidths, it offers a significant improvement over existing methods.
Under this proposed association policy, we derive the association probability, coverage probability, and spectral efficiency. 
Through both analysis and simulation, we assess the proposed association policy over a traditional MAP association policy, which confirms it as a simple yet effective route to boost UAV network performance. 
Future work may explore other factors to optimize network performance.

\bibliographystyle{IEEEtran}
\bibliography{IEEEabrv, UAV}  
\end{document}